\documentclass[preprint]{aastex}

\begin{document}

\title{Phase Drifts of Sub-pulses during the 2004 Giant Flare of SGR 1806-20 and Settling of the Magnetic Fields}
\author{Yi XING and Wenfei YU}
\affil{Key Laboratory for Research in Galaxies and Cosmology and Shanghai Astronomical Observatory, 80 Nandan Road, Shanghai 200030, China. E-mail: wenfei@shao.ac.cn}

\begin{abstract}
We analyzed the observations of SGR 1806-20 performed with the \textit{Rossi X-ray Timing Explorer} (RXTE) during its 2004 giant flare. We studied the phase evolution of the sub-pulses identified in the X-ray waveform and found that the sub-pulses varied in phase with time and then gradually settled, which might indicate drifts of the emission regions in relative to the neutron star surface, or changes in the local emission geometry before the magnetic field became stable. The characteristic e-folding timescale of the phase drifts measured starting about 15 s following the initial flux spike are in the range between 37 s and 84 s. This leads to the first measurements of the characteristic timescale for the magnetic field of the neutron star to settle after a field reconfiguration during the giant flare. 
\end{abstract}

\keywords{neutron stars, Soft Gamma-ray Repeaters}

\section{Introduction}
Soft Gamma-Repeaters (SGRs) are thought to be a type of magnetars: isolated neutron stars with much higher magnetic field than normal pulsars. The characteristic strength of magnetic field is believed to be about $10^{14}-10^{15}$ G, inferred from the observational measurements of the pulse period and its derivative assuming a dipole field. Recurrent X-ray and soft gamma-ray flares have been seen from SGRs, with a typical luminosity up to $10^{42}$ ergs/s since their discovery in 1979 (for an observational  review, see Mereghetti 2008). Several major observational properties of the flares in SGRs can be explained by the most popular model in which the flares are produced by Alfv\'{e}n wave which propagates in the magnetosphere of the neutron star, due to the magnetic field diffusion associated with small cracks in the neutron star crust \citep{tho95}. Besides these recurrent flares, several giant flares with much higher luminosity have been observed. The luminosity during giant flares could reach more than a million times the Eddington luminosity of a common neutron star, i.e. $10^{44}-10^{46}$ erg/s. As an example, for the most extreme giant flare observed from SGR 1806-20 in 2004, the peak luminosity was about $2\times10^{46}$ erg/s \citep{hur05,mer05,pal05,ter05}. Energy of giant flares was thought to have a magnetospheric origin other than a neutron star crust origin. In the above magnetar model, giant flares are caused by a fireball of pair-dominated plasma that expands at relativistic speed.  The unwinding internal magnetic field will build up the magnetic energy outside the neutron star slowly by pushing out inside electric currents into the magnetosphere. When the magnetosphere currents reach a point at which the magnetosphere becomes dynamical unstable, the twist field outside the star associated with dissipation will have a sudden relaxation and the magnetic field will have a reconfiguration, consequently the destructive magnetic instability may produce the neutron star crust fracturing in large scale, which can lead to an extreme outburst \citep{tho95,lyu06}.

The total energy inferred from the magnetic strength of magnetar is of about $10^{47}$ erg, which is significantly more than energy contained in common neutron stars in the form of rotational kinetic energy \citep{bra06}. During a giant flare a large portion of the energy contained in magnetar would be lost, so the giant flares are very rarely seen. Up to now only three events have been detected: the 1979 flare from SGR 0526-66, the 1998 flare from SGR 1900+14, and the 2004 flare from SGR 1806-20. Interestingly, it was suggested before the 2004 Giant flare that magnetars would have a supergiant flare produced by a dynamic overturning instability. These type of flare should be 100 times less frequent than normal giant flares but could be observed within about 10 times the distance because of its much higher energy release \citep{eic02}. The 2004 giant flare was just like the predicted events.

Giant flares have some common properties. In the X-ray waveband the light curves of giant flares consist of a short hard initial spike followed by a long softer pulsating tail. The peak luminosities of the 1979 flare and the 1998 flare were about $10^{44}$ erg/s \citep{maz79,fen81,hur99,fer99,fer01} while during the 2004 flare it reached to $2\times10^{46}$ erg/s \citep{hur05,mer05,pal05,ter05}. The rise time is typically smaller than a few milliseconds during which the luminosities of giant flares increase rapidly. After the initial spike there is the pulsating tail lasting about hundreds of seconds, which is thought to originate from the cooling of the ejected plasma which remains confined in a trapped thermal fireball near the stellar surface due to the strong magnetic field in magnetar model \citep{tho95}.

For the three known giant flares, the pulsating tails always show a strong periodic modulation at the neutron star spin period and have complex pulse profiles. They are usually not single-peaked and show strong evolution with time. During the 1998 flare from SGR 1900+14, the pulse profile of the pulsating tail showed four peaks and a dip which were almost evenly spaced at about 1.03 s interval in each period of 5.16 s \citep{fer01}, and the four-peak structure evolved to being nearly sinusoidal with time. The multi-peak features are thought to be described as collimated X-ray jets from trapped fireballs, in comparison to the 1979 giant flare from SGR 0526-66 during which only two sub-pulses have been observed. It was suggested that the number of sub-pulses detected in each period of these two sources could be due to different number of X-ray jets \citep{tho01}. For the 1998 giant flare the phases of these four sub-pulse peaks were recognized stable, which may suggest that the X-ray jets are tied to surface features on the neutron star. The four-peaked pattern is thought to be direct evidence of multipolar structure of the neutron star's magnetic field \citep{fer01}. Investigations in the quiescent emission of SGR 1900+14 revealed that before the 1998 giant flare the pulse profile of the source always showed complex multi-structure \citep{woo01}, and once it evolved to sinusoidal during the flare it keeped unchanged for more than 1.5 years. It is suggested that if the multipeak profile indicates multipolar field structure and the simple sinusoidal profile indicates dipolar field geometry, during the time of the 1998 giant flare the magnetic field, both inside and outside the neutron star, underwent a global reconfiguration, while during the quiescent stage the magnetic field was relative stable \citep{woo01}.

The evolution of pulse profile was also observed during the 2004 flare from the SGR 1806-20. It showed an opposite trend compared with the 1998 flare; the pulse profile evolved from a simple sinusoidal to a complex multi-peaked structure \citep{woo06}. The phases of these multi-peak are generally thought stable, indicating that there was no substantially field change during the tail \citep{pal05,mer05,boggs07}, even though we see obvious phase shifts of the sub-pulses on short time scales (see our study below). In the quiescent X-ray light curve prelude to the 2004 flare, the pulse profile was nearly simple sinusoidal, it became more complex after the flare. Considering the pulse profile changes during the flare, it was recognized that the pulse profile changes in the quiescent light curve were flare-induced, consistent with SGR 1900+14 \citep{woo07}.

Besides the two SGRs from which evolving pulse profile has been observed, during magnetar flares from magnetar candidates, for example CXOU J164710.2-455216, the pulse profile was also observed to evolve. During the observations of 4.3 day prior to and 1.5 day subsequent to two remarkable events that were detected with Swift on 2006 September 21 (a 20-ms burst and a glitch), the pulse profile was found to change from single-peaked to three distinct peaks \citep{mun07}. For this type of source with lower X-ray energy release, a large-scale rearrangement of the magnetic field was thought unreasonable. It is suggested that the pulse shape was governed by the distribution of currents within the magnetosphere, and the profile change should be due to the change in the distribution of currents in the magnetosphere \citep{mun07}.

In the following study we focus on the pulse profile during the pulsating tail of the 2004 flare form SGR 1806-20. In our previous study of SGR 1806-20 giant flare using {\it GoodXenon} mode data of the RXTE/PCA {\it only}, we have decomposed this pulse profile into four sub-pulses. We modeled the pulse profile with four Gaussian functions and a constant background flux level, and investigated the X-ray waveform evolution of each sub-pulses, excluding the first 2-3 cycles because of data gaps in the {\it GoodXenon} mode data \citep{xin09}. In that study we found the phases of the sub-pulses were not fixed during the tail. The phase of one sub-pulse was decreasing while the phase of another sub-pulse was increasing with time -- an apparent anti-correlation between the phase drifts of the two sub-pulses. This suggests that the phases of the sub-pulses moved although the multi-peak waveform structure remained unchanged. Modeling the sub-pulses as multi-peak Gaussian gives good description of the overall phase distribution of the photons in each sub-pulses, but not accurate and sensitive in determination of the phases of the sub-pulse peaks. Furthermore, the data gap in the {\it GoodXenon} mode data recorded at the beginning of the Giant flare because of telemetry saturation prevents us from studying the sub-pulse phase evolution during the first few cycles of the giant flare. In this study, we intend to measure the phases of the sub-pulse flux peaks, which show the sub-pulse phase shifts more obviously. We determined the sub-pulse phase by 7-point Gaussian fit to each sub-pulse peak with {\it Standard 1} mode data instead of {\it GoodXenon} data. This data mode does not have data gaps and uniquely helps determination of the sub-pulse phases during the first few cycles. We found strong evidence of systematic phase shifts of those sub-pulse peaks, in agreement with our previous results but further helped us measure the characteristic timescales of the phase evolution of the sub-pulses. This timescale might indicate the timescale for the magnetic field to settle on the neutron star after a reconfiguration, which has been suggested to associate with the giant energy release during the flare. 

\section{Observations and data analysis}
The giant flare of SGR 1806-20 was recorded by the \textit{Rossi X-ray Timing Explorer} (RXTE) on 2004 December 27th, during that time the RXTE was pointing at the cluster of galaxies A2163 (with the RXTE Observation ID: 90132-01-13-09). The data we used was taken by Proportional Counter Array (PCA) in {\it GoodXenon} mode with a time resolution of 9.5367431640625e-07 s (for determining pulse period and checking if the sub-pulse phase drift was not obviously energy dependent) and {\it Standard-1} mode data with a time resolution of 0.125 s (for investigating sub-pulse phase drift).

We used the Standard-1 mode data to generate the light curve of SGR 1806-20 during the giant flare in  Figure 1. The time resolution is 0.125 s. The X-ray emission during the giant flare consists of two components: a descending non-pulsed component and a pulsed component. In order to accurately determine pulse period and sub-pulse phase during the tail, we need to remove the non-pulsed component from the light curve. This was not done in our previous work \citep{xin09}. We separated the light curve into segments of cycles starting at the first period during the giant flare (the phase-zero was selected at the start of the 3rd period, corresponding to the RXTE raw space craft clock time of 346800645.87843037 s) using a rough period of 7.56 s as reported \citep{isr05} and calculated the average count rates for each cycle. This is shown in Figure 1. 3-points quadratic interpolation was applied to the average count rates in order to construct the non-pulsed component. By removing the non-pulsed component from the high time resolution light curve (1/4096 s, generated from GoodXenon mode data), we obtained the pulsed component during the pulsating tail, which was used to determine the pulsed component and the sub-pulse phase shift. Considering the rough period was determined in a way influenced by significant profile changes during the first few cycles, we determined the pulse period during a 400s interval starting at about 49 seconds after the initial spike when the pulse profile was relatively stable. We folded the 400 s data into 128 phase bins and searched for the best period in the period range from 7.5536 s to 7.5663 s using a step-length of 0.0001 s. A best period of 7.5611 s was obtained, which is different from the 7.5605 s period determined in our previous approach without removing the non-pulsed component. Using this period, we generated pulse waveform of each cycle during the pulsating tail. We also checked the waveforms in the two energy ranges in which the total counts were nearly equal, i.e. 2-15 keV and 15-60 keV. We concluded that the waveform during the pulsating tail did not show obvious energy-dependent phase shift. Therefore we went further using the light curve in the entire PCA band to study the sub-pulse evolution.

We then apply the newly determined pulse period to the {\it Standard-1} mode data to study the waveform evolution during the pulsating tail, because the data covers the gaps in the GoodXenon mode data at the beginning of the giant flare introduced by the saturation of RXTE telemetry while the time resolution is good enough for sampling on the time scale of the pulse period. Following the previous procedure to remove the non-pulsed component, we obtained the pulsed component during the pulsating tail using the period of 7.5611 s. Then we plotted the waveforms of the pulsed component of each cycle in Figure 2. A total of 35 cycles were shown. A portion of the inferred pulsed emission during the first cycle was invisible (dashed line) because the non-pulsed component was overestimated assuming a linear interpolation. From Figure 2 we could see that the overestimation was introduced by the influence of the initial spike in the first cycle. 

Multi-Peak structures could be clearly seen from these pulse profiles, as shown in Figure 2. Each pulse profile consists of four sub-pulses, and the phases of the sub-pulses were not fixed during the flare. A circle was used to mark the peak value of the major sub-pulse of the first cycle, which seemed to relate to the 3rd sub-pulse. This indicates that the waveform evolved drastically at the beginning of the giant flare.  In order to determine the phases of the sub-pulses, we selected seven data points centered at the peak of each sub-pulse and fitted them with a single Gaussian function. Notice that these pulse profiles were obtained by removing the trend of the non-pulsed component and the mean count rate level during the cycle has been included. We found in most cases the sub-pulse peaks could be fitted by Gaussian functions with reduced $\chi^2$ in the range between 0.20 and 9.4. As an example, the pulsed profile of the 8th cycle and the Gaussian components resulted from those seven-points Gaussian fits are shown in Figure 3.

\section{Results}
\subsection{Phase drifts of sub-pulses}
We show the evolution of the phases of the sub-pulses in Figure 4. We called the sub-pulse at about phase 0.2 as the 1st sub-pulse, the sub-pulse at about phase 0.45 as the 2nd sub-pulse, the sub-pulse at about 0.55-0.6 as the 3rd sub-pulse, and the sub-pulse at about 0.73 as the 4th sub-pulse. The 3rd sub-pulse was excluded in our analysis because it overlaps with the 2nd sub-pulse and  its phase can not be well-determined. In addition the  reduced chi-square values were greater than 10 in the Gaussian fits for a few cycles. The large reduced chi-squares indicated that the peaks of these sub-pulses can not be modeled by single Gaussian functions. Therefore we did not measure the phases of these sub-pulses, respectively. These cycles are: the first two cycles, the 16th, the 17th, and the 34th cycles for the 1st sub-pulse; the first two cycles for the 2nd sub-pulse; and the 1st, the 16th, the 20th, and the 31th cycles for the 4th sub-pulse.

Figure 4 shows clearly that the phases of the sub-pulses were not fixed. The evolution trend of the 1st sub-pulse and the 4th sub-pulse is opposite, especially in the early stage where the phase of the first sub-pulse decreases while the the phase of the fourth sub-pulse increases. We calculated the correlation coefficient between these two sub-pulses and got -0.72$\pm$0.12, confirming that the two sub-pulse phase drifts are opposite. As suggested previously in \citep{xin09}, the correlation may suggest the phase drifts are physically connected. For example, the two sub-pulses may be associated with opposite magnetic poles since their phase separation is around 0.5. 

\subsection{Characteristic timescales of phase drifts}
We have measured the phase evolution of the 1st sub-pulse since the third cycle and the 4th sub-pulse since the second cycle with exponential models to determine the e-folding timescales for the phases to become stable. The e-folding time scales are 36.95$\pm$3.01 s for the 1st sub-pulse and 65.41$\pm$5.99 s for the 4th sub-pulse (see Figure 5). In order to investigate whether smaller e-folding timescales would be obtained if we include cycles at earlier stages, which would suggest that our results be biased because of lack of data during the first two cycles, we investigated if smaller e-folding times would be obtained if we make use of data from earlier cycles. By including cycles from the 10th and backward one by one, we checked if the e-folding times obtained from the fits would get smaller. We found in general these e-folding times are within uncertainties. Therefore the characteristic time scales of tens of seconds we obtained indeed represents the time scale for the phases of the sub-pulses to become stable. 

The phase separation of the 2nd sub-pulse and the 3rd sub-pulse seemed more obviously changing in the early stage too (see Figure 2), while in the later stage their phases seemed relative fixed. The phase of the 2nd sub-pulse showed no gradual change while the phase of the 3rd sub-pulse modeled with Gaussian components in our previous preliminary study was increasing with time. Adopting the method used in \citep{xin09} by decomposing the waveform into four Gaussian components, but without a constant component since the long-term background trend was removed, we obtained the phase evolution of the 3rd sub-pulse (note: the method is different from those for the 1st and the 4th sub-pulses). We determined the e-folding timescale as 84.17$\pm$7.54 s, on which the phase became stable (also see Figure 5). The three best-fit exponential models were over-plotted as thick lines in Figure 2 . 

From Figure 2 we could see sub-pulse 1 and 4 are sitting on either sides of sub-pulse 2. Flux contribution from sub-pulse 2 may affects our estimates of the phases of sub-pulses 1 and 4, which tends to give sub-pulse phases toward sub-pulse 2. In order to investigate this, we also included data points of the entire sub-pulses rather than only 7 data points of sub-pulses 1 and 4 as well as the data of the slope on either side of sub-pulse 2. For sub-pulse 4 the minimum phase chosen corresponds to the flux minimum during the rising side of the sub-pulse, which was in the phase range between 0.5 and 0.7, and the maximum phase chosen was phase 1.0. For sub-pulse 1 the 7 data points we selected before and a background data slope covers the entire sub-pulse, i.e., we included data points between phase 0.0 and maximum phase of the 7 data points. We then fitted the data associated with sub-pulses 1 and 4 to a Gaussian function plus a linear component, representing the sub-pulse peak and the sloping side of sub-pulse 2, respectively. The slope of the linear component was set to be above 0 for sub-pulse 1 and to be below 0 for sub-pulse 4, as sub-pulse 1 sits on the left shoulder of sub-pulse 2, and sub-pulse 4 sits on the right shoulder. An example of the fits for the 3rd cycle of sub-pulse 1 and 4 is shown in Figure 6 and 7. We then obtained the phase evolution of the sub-pulses. Similarily we ignore those fits with reduced-$\chi^2$ larger than 10. We then fitted the obtained phase evolution with the exponential model to measure the e-folding timescales for the phases of the sub-pulses to settle and got 34.05 +/- 3.07 s for sub-pulse 1 and 50.94 +/- 6.98 s for sub-pulse 4 (see Figure 8). This indicates that the e-folding time scales are indeed on the time scales of tens of seconds.

\subsection{Comparison to the decay timescale of the giant flare}
The e-folding decay time of the 1998 giant flare from SGR 1900+14 was found to vary with energy, e.g., 78 s in the Ulysses energy range of 40-100 keV and 70 s in the BeppoSAX energy range of 100-700 keV \citep{fer01}. An even longer decay timescale of 90 s \citep{maz99} was obtained in the Konus energy range. It was therefore inferred that the e-folding decay time of the 1998 flare decrease with increasing photon energy \citep{fer01}. For the 2004 giant flare from SGR 1806-20, an e-folding decay timescale of 128.21$\pm$0.19 was obtained in the RXTE/PCA energy range (2--60 keV). Therefore, the e-folding decay time of the X-ray flux of the giant flare is a little larger than the characteristic timescale for the phases of the sub-pulses to settle, but on the same order of magnitude.

\section{Conclusion and discussion}
We have determined the phases of the sub-pulses in the X-ray waveform during the 2004 giant flare and studied their evolutions. There existed phase drifts of these sub-pulses up to 0.06 in less than 100 seconds. The phases of the sub-pulses became stable after a characteristic timescale of tens of seconds. This indicates that the magnetic instability responsible for the magnetic field reconfiguration damps out on the above timescale and the magnetic field became stable after such a characteristic time scale. Two sub-pulses separated by about 0.5 phase, namely the 1st sub-pulse and the 4th sub-pulse in our study,  show opposite evolution trend. The apparently correlated evolution trend is more obvious during the early stage of the giant flare. If the two sub-pulses correspond to the emission regions of two opposite magnetic poles or of the same magnetic pole, the phase evolution and the opposite phase trend indicate the two emission regions are probably tied together during the giant flare. The $\sim0.5$ phase separation of the two sub-pulses and the potentially physically correlated phase drifts then likely provide evidence that we can see at least two poles of the neutron star. This provides an additional evidence for a multipolar magnetic field geometry other than that discussed in Feroci et al (2001) on the four sub-pulse profile.

The initial spike of a giant flare has a sharp rise which is typically smaller than a few milliseconds. For the 1979 event the rise time was smaller than 2 ms. For the 1998 event it was smaller than 4 ms \citep{maz99}. During the 2004 giant flare from SGR 1806-20 the rise time was much smaller than ever observed, which took only about 0.3 ms \citep{pal05}. The steep rise time of giant flare is thought associated with the Alfv\'{e}n crossing time of the inner magnetosphere, which is as short as $30 \mu s$ estimated by $R_{NS}/c$, where $R_{NS}$ is the radius of the neutron star and $c$ is the light speed \citep{lyu06,tan07}. During the steep rise the twist field outside the neutron star suffers a sudden relaxation, which leads to a rapid release of energy stored in the magnetosphere \citep{lyu06,bra06}. Besides the initial steep rise there was another timescale noted in previous studies, i.e. the intermediate rise time to the peak. The intermediate rise time to the peak during the giant flare from SGR 1900+14 was 3.1 ms, which was shorter than the 9.4 ms observed during the 2004 flare from SGR 1806-20. This timescale is thought to be limited by the propagation of a fracture, so it could be used to estimate the fracture size during giant flares \citep{tan07}.

For all of the three giant flares observed, the duration of the initial spike was about hundreds of milliseconds: 0.25 s for the 1979 giant flare from SGR 0526-66; 0.35 s for the 1998 giant flare from SGR 1900+14; and 0.5 s for the 2004 giant flare from SGR 1806-20 \citep{mer051}. It may reflect the Alfv\'{e}n wave crossing time of the star which is in the range between 0.2 s and 0.5 s estimated by $R_{NS}/V_{A,NS}$, where $V_{A,NS}$ is the Alfv\'{e}n wave crossing speed in the star \citep{tho95}.

The e-folding decay time of the giant flare was thought to describe the cooling time of the trapped fireball, which was formed by energy released in the initial spike trapped in the neutron star magnetosphere by high magnetic field \citep{tho95,pal05,mer08}. The emission from trapped fireball in higher energy band seemed to cool a little faster than that in lower energy band, which leads to a softening of the dominated emission during the later part of the pulsating tail. The characteristic time scales of the phase drifts of the sub-pulses are a little smaller, but comparable to the decay time scale of the giant flares. We therefore infer that the magnetic field reconfiguration settle earlier than the cooling of the trapped fireball in the magnetar model. 

In this work we have found the phases of the sub-pulses were obviously drifting during the first ten cycles or so and gradually became stable. The characteristic timescale of phase drifts is in the range 37 s -- 84 s. This may tell us the instability of the magnetic field during the giant flare damped in tens of seconds before the field configuration became stable. Thus this time scale may be used to constrain the magnetic fields and the mechanism for the energy release of the giant flare. 

\begin{figure}
\includegraphics[width=6in]{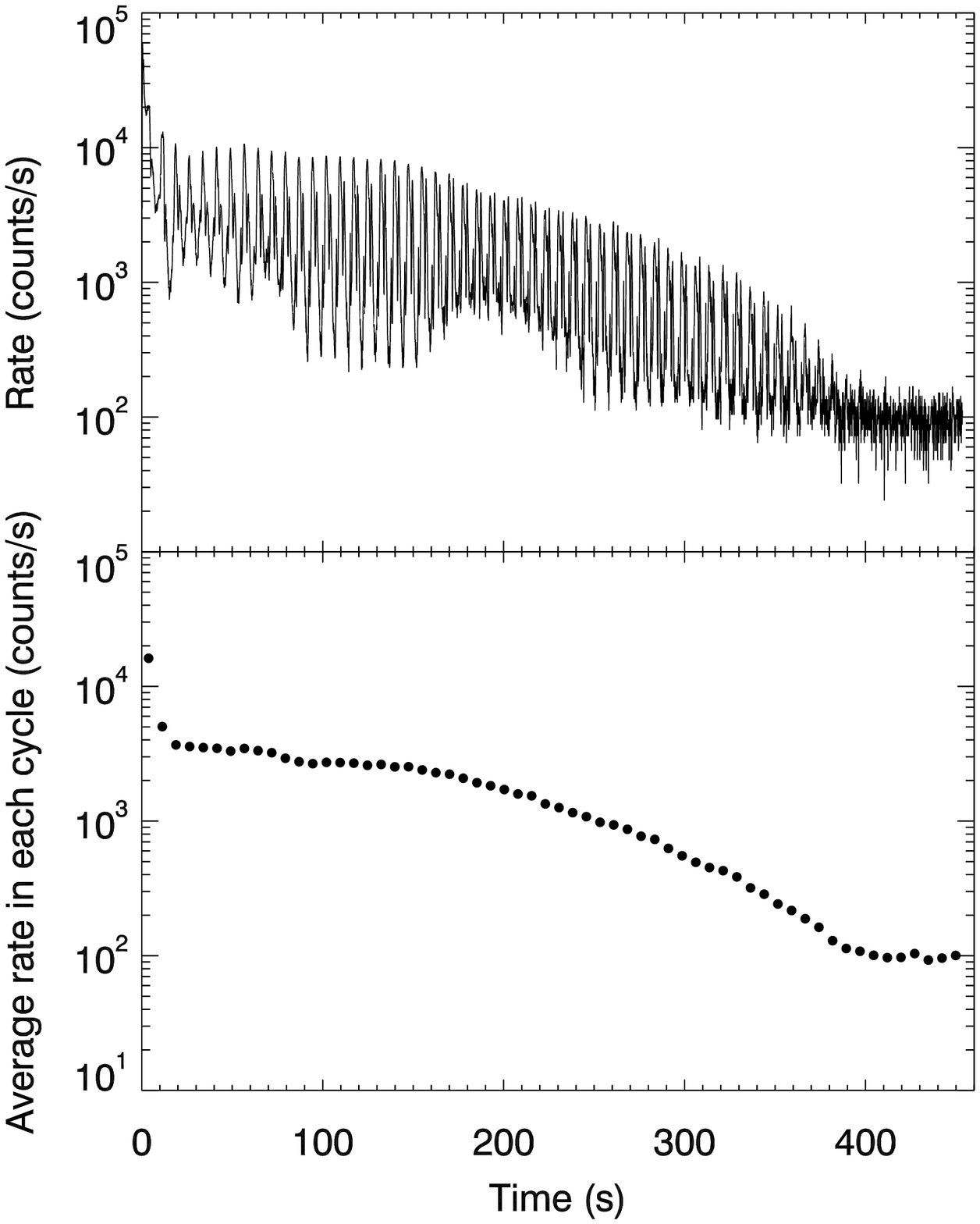}
\caption{Top: the (2-60 keV) RXTE/PCA light curve during the 2004 outburst of SGR 1806-20 generated from the Standard-1 mode data. The time resolution of the light curve is 0.125 s. Bottom: evolution of the average rates of each cycles during the pulsating tail.}
\end{figure}

\begin{figure}
\includegraphics[width=6in]{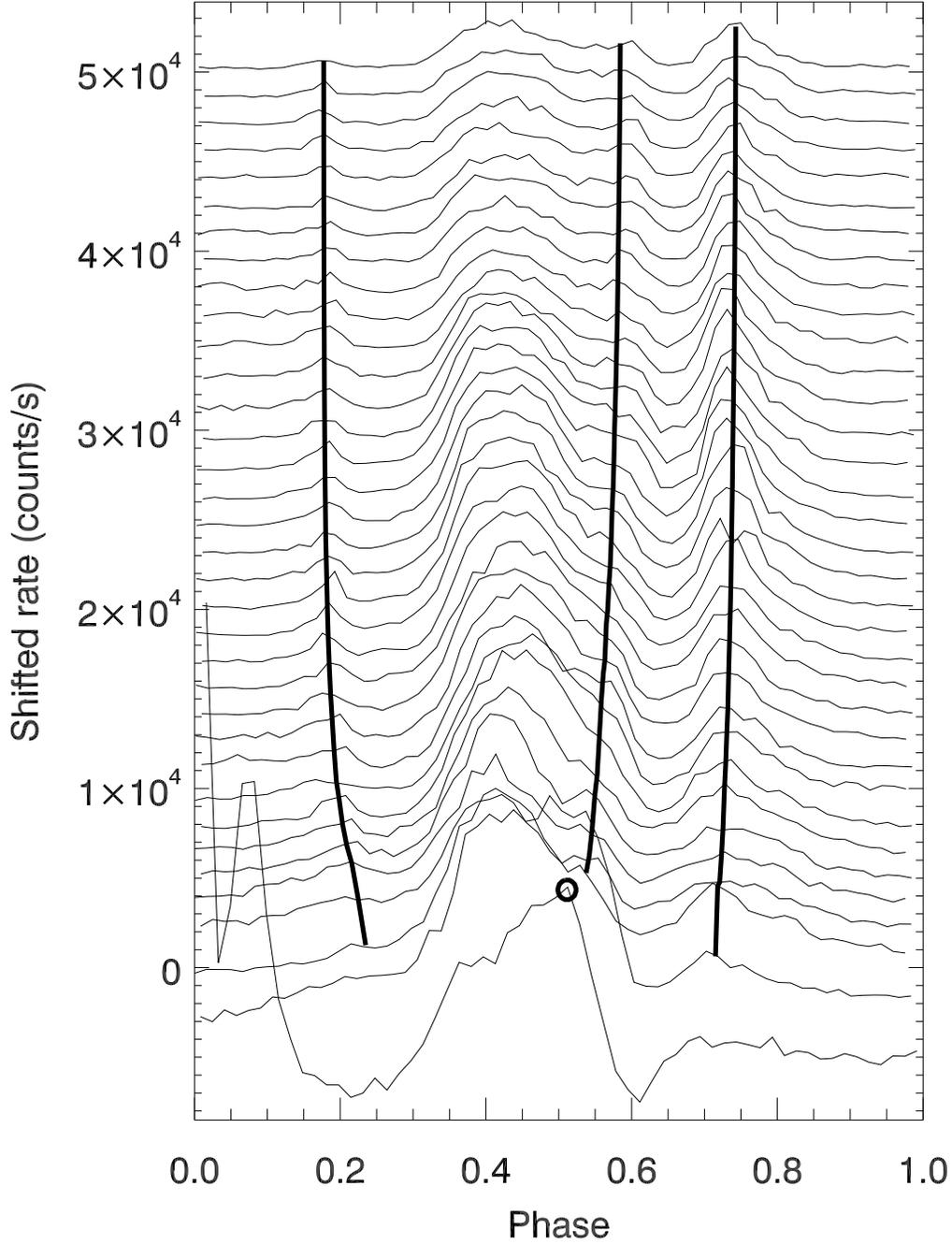}
\caption{Pulse profile evolution during the first 35 cycles of the 2004 giant flare obtained from {\it Standard 1} data. The non-pulsed background trend is removed from the data. Time sequence increases from bottom to top. Count rate corresponding to each cycle was shifted by 1500 counts/s one by one. The circle marks the peak of the major sub-pulse during the first period, consistent with the sub-pulse \#3 identified in subsequent cycles.}
\end{figure}


\begin{figure}
\includegraphics[width=6in]{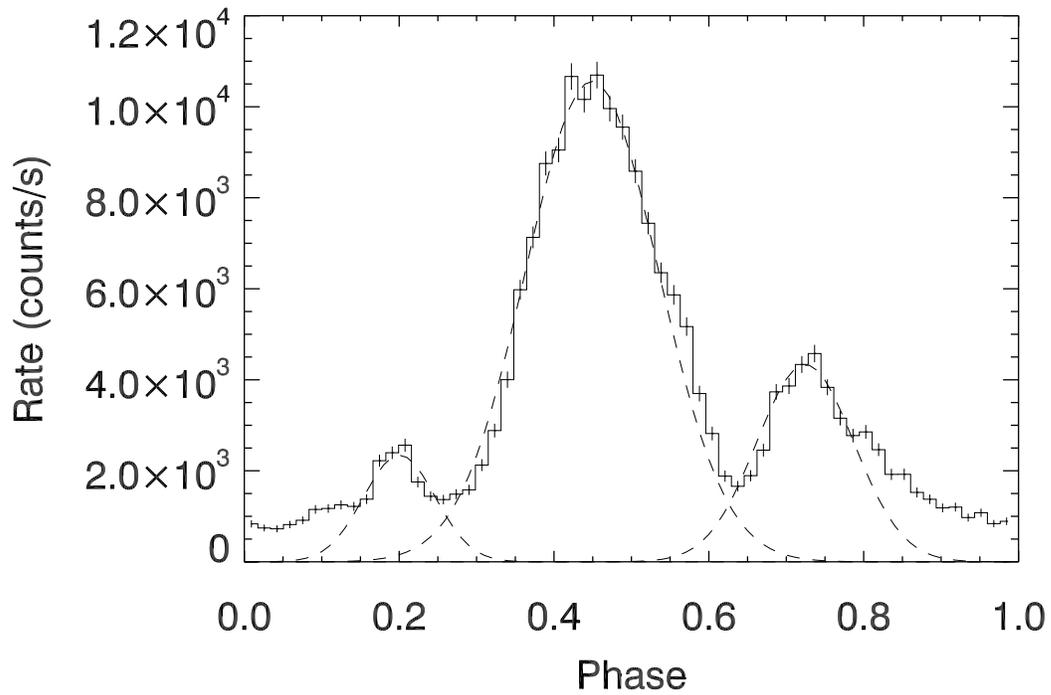}
\caption{An example of the model fit to the waveform in which each of the three major sub-pulse peaks were fitted with a Gaussian (seven-point fits). The best-fit Gaussians were over-plotted as dashed lines. The central phases of the Gaussians gives the phases of the sub-pulses. A constant level at the mean count rate of the cycle has been added.}
\end{figure}

\begin{figure}
\includegraphics[width=6in]{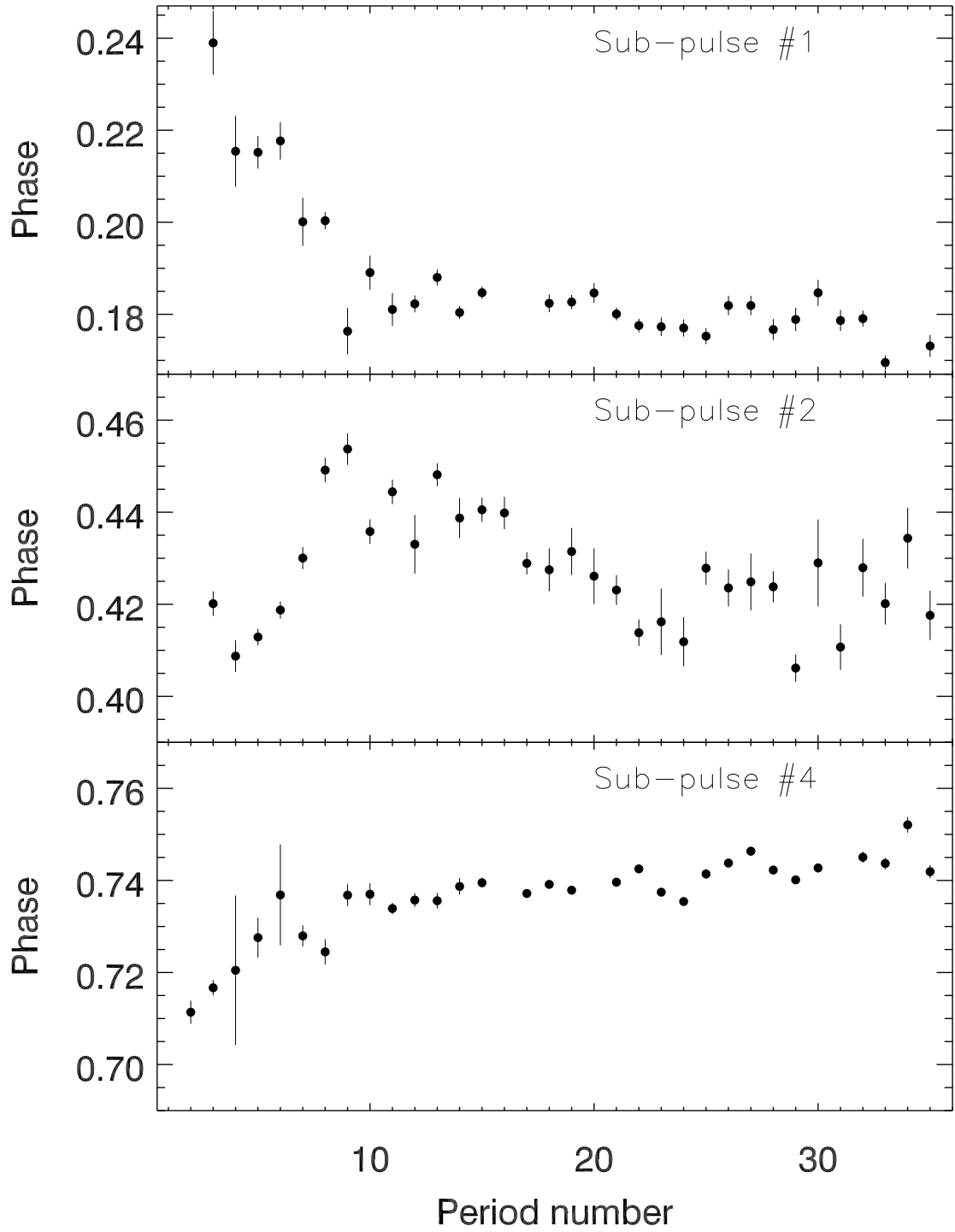}
\caption{Phase evolution of the sub-pulses \#1, \#2, and \#4 measured by seven-points Gaussian fits to the sub-pulse peaks.}
\end{figure}

\begin{figure}
\includegraphics[width=6in]{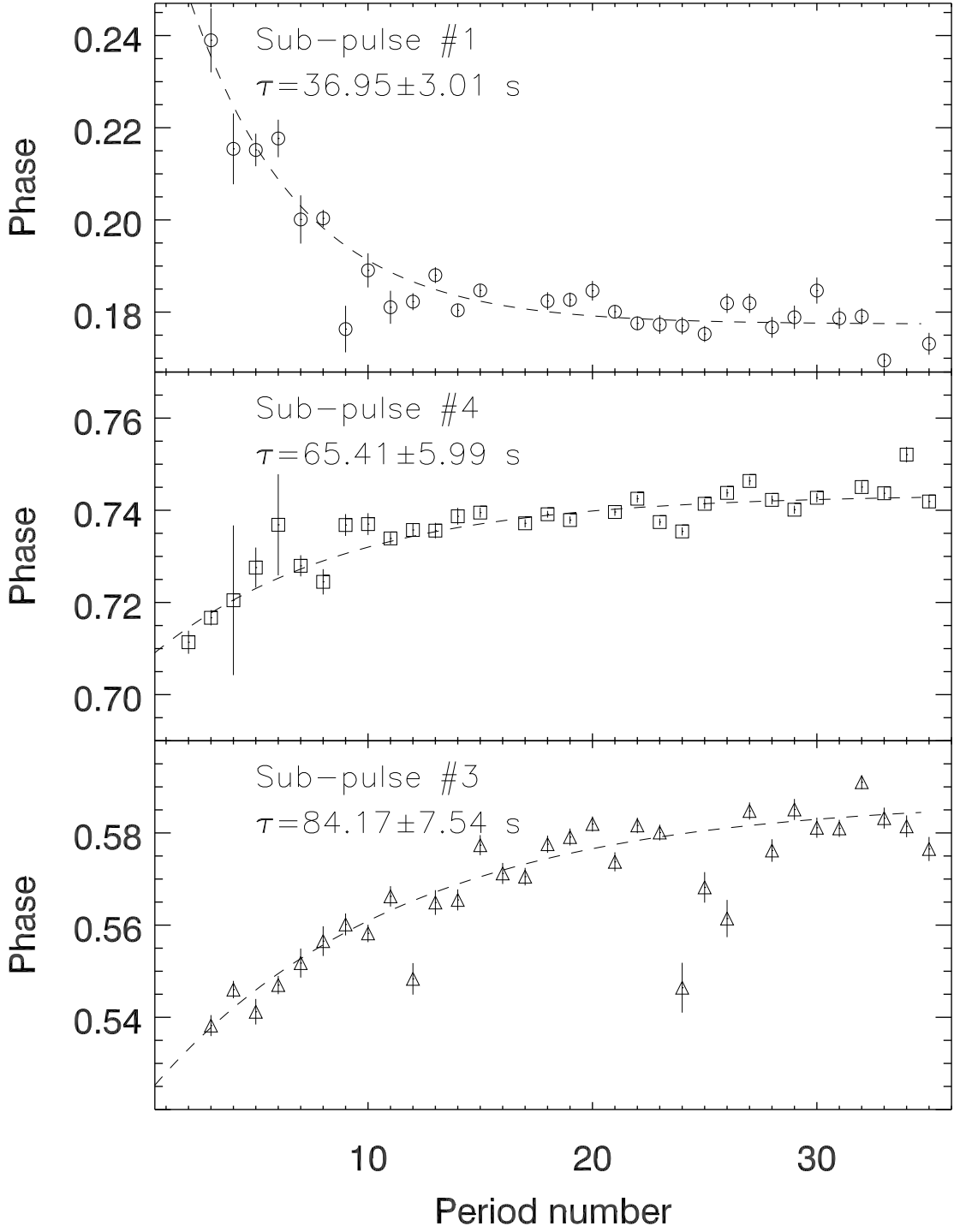}
\caption{Phase evolution of the sub-pulses modeled with exponential phase drift. Notice that the phases of sub-pulse \#3 was obtained from a decomposition of the entire waveform into 4 Gaussian components and the phases of the \#1 and \#4 were obtained from 7-points Gaussian fits. The phase drifts should be considered as those in relative to the sub-pulse \#2 -- the major sub-pulse.}
\end{figure}

\begin{figure}
\includegraphics[width=6in]{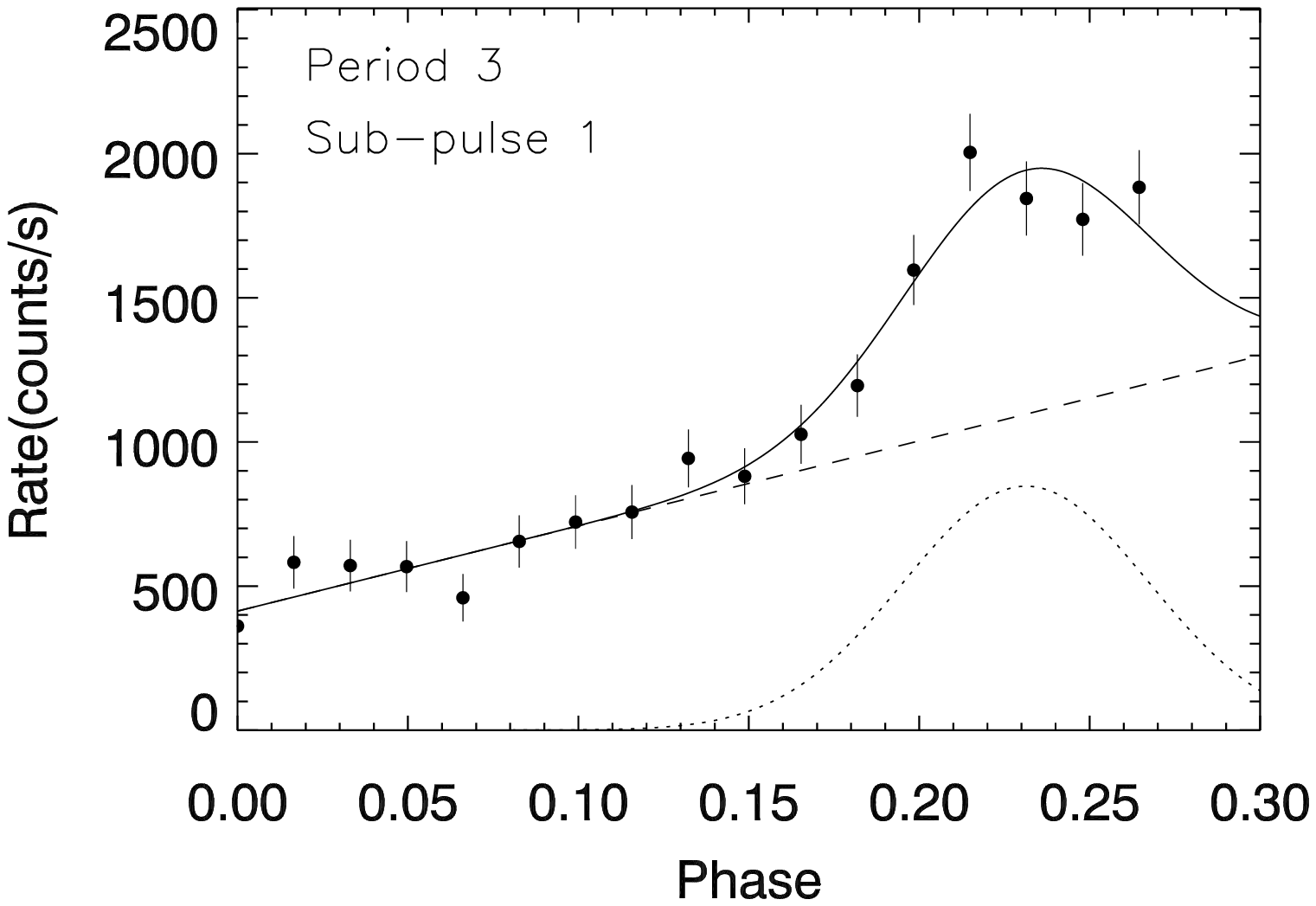}
\caption{The model fit to data around sub-pulse 1 considering the left shoulder of sub-pulse 2 of the 3rd period. }
\end{figure}

\begin{figure}
\includegraphics[width=6in]{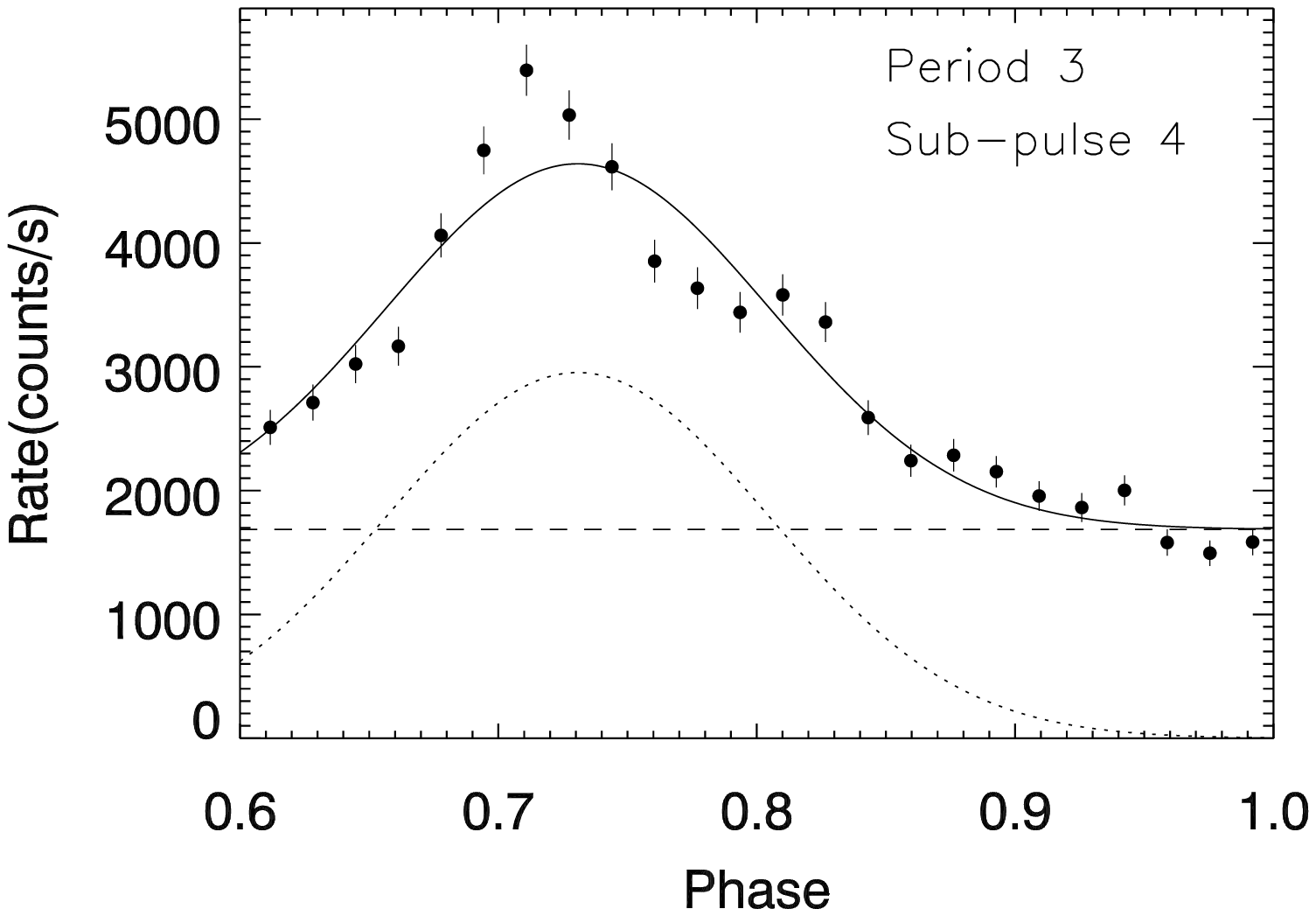}
\caption{The model fit to data around sub-pulse 4 considering the right shoulder of sub-pulse 2 of the 3rd period. }
\end{figure}

\begin{figure}
\includegraphics[width=6in]{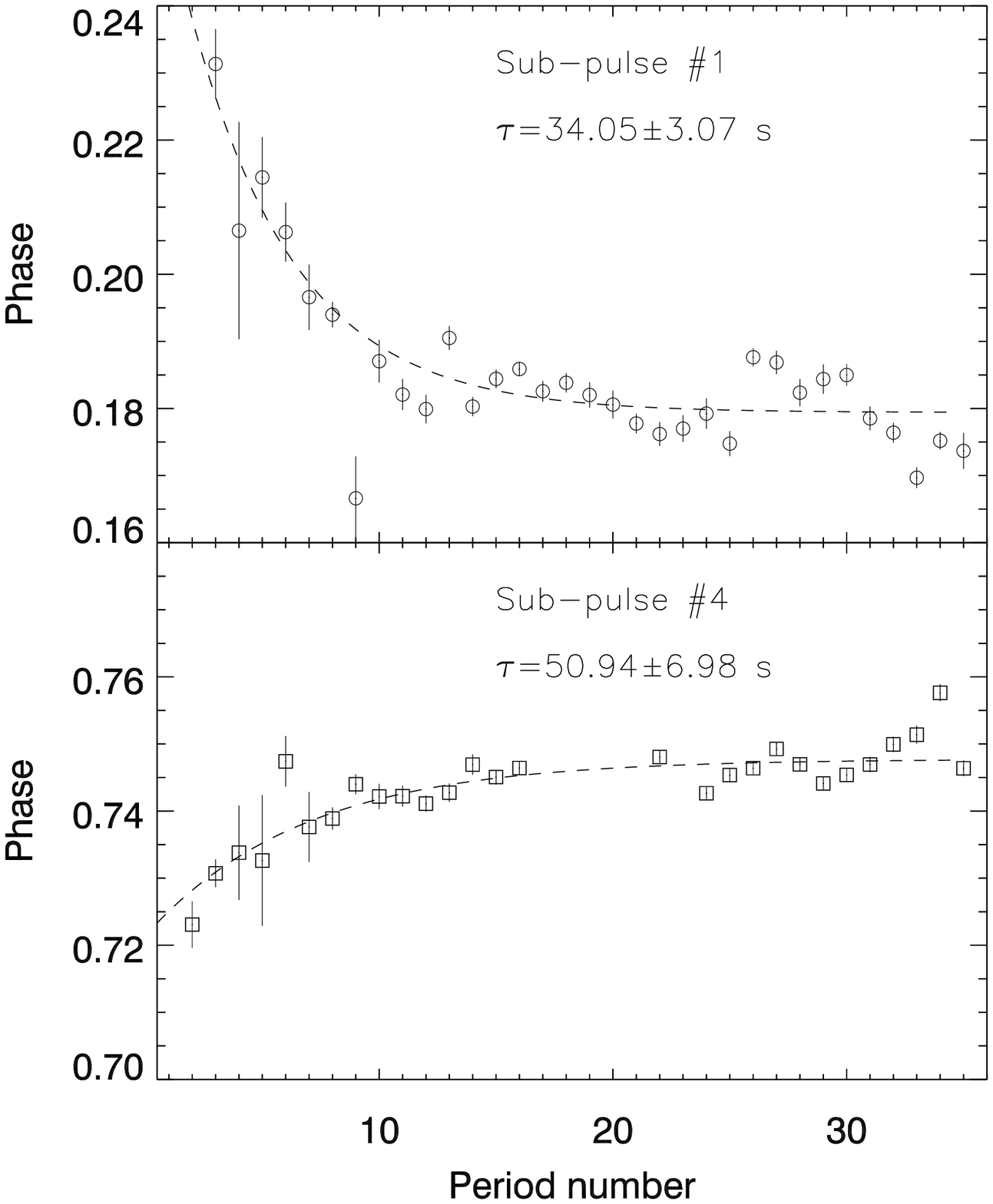}
\caption{Phase evolution of the sub-pulses modeled with exponential phase drift. Different from Figure 5, the phases of the sub-pulses 1 and 4 were obtained from fitting with the sloping sides of sub-pulse 2 to a Gaussian function plus a linear slope. Notice that data with reduced-$\chi^2$ larger than 10 were excluded in the plot. Around period  No. 17--23 the model does not fit sub-pulse 4 well. }
\end{figure}

\acknowledgments
WY would like to thank Dong Lai of Cornell University, Chryssa Kouveliotou and Peter Woods of  the National Space Science and Technology Center (NSSTC), and Tod Strohmayer of NASA/Goddard Space Flight Center of the US, for useful comments and encouragement. We also thank the anonymous referee for a careful reading of the manuscript and for his/her useful comments and suggestions on estimating the contribution to the QPO frequency width from drifting emission regions on the neuron star. This work was supported in part by the National Natural Science Foundation of China (10773023, 10833002, 11073043), the One Hundred Talents project and the Pre-phase Studies of Space Science Projects of the Chinese Academy of Sciences, the Shanghai Pujiang Program (08PJ14111), the National Basic Research Program of China (2009CB824800), and the starting funds of the Shanghai Astronomical Observatory. The study has made use of data obtained through the High Energy Astrophysics Science Archive Research Center Online Service, provided by the NASA/Goddard Space Flight Center.

\end{document}